\documentclass[runningheads]{llncs}

\usepackage{lipsum}

\newcommand\blfootnote[1]{%
  \begingroup
  \renewcommand\thefootnote{}\footnote{#1}%
  \addtocounter{footnote}{-1}%
  \endgroup
}
\usepackage{graphicx}
\usepackage{hyperref}
\usepackage{multirow}
\usepackage[normalem]{ulem}
\hypersetup{
    colorlinks=true,
    linkcolor=blue,
    filecolor=magenta,      
    urlcolor=cyan,
}
\usepackage{amsmath}
\usepackage{amsfonts}

\DeclareMathOperator*{\argmin}{arg\,min}
\newcommand{\norm}[1]{\left\lVert#1\right\rVert}

\usepackage{tikz}
\usepackage{mathtools}
\usetikzlibrary{shapes,arrows}
\usetikzlibrary{positioning}
\tikzstyle{decision} = [diamond, draw, fill=blue!20, 
    text width=4.5em, text badly centered, node distance=3cm, inner sep=0pt]
\tikzstyle{block} = [rectangle, draw, fill=gray!20, 
    text width=3em, text centered, rounded corners, minimum height=4em]
    \tikzstyle{dblock} = [rectangle, draw, dashed, fill=blue!20, 
    text width=3em, text centered, rounded corners, minimum height=4em]
    \tikzstyle{rblock} = [rectangle, draw, fill=blue!20, 
    text width=3em, text centered, rounded corners, minimum height=4em]
\tikzstyle{line} = [draw, -latex']
\tikzstyle{cloud} = [draw, ellipse,fill=red!20, node distance=3cm,
    minimum height=2em]
\tikzstyle{blank} = [node distance=3cm,
    minimum height=2em]   
    
\begin{document}

\title{Neural Network-based Reconstruction in Compressed Sensing MRI Without Fully-sampled Training Data}
\titlerunning{Unsupervised Deep Reconstruction of CS-MRI}
%
\author{Alan Q. Wang\inst{1} \and
Adrian V. Dalca \inst{2, 3} \and
Mert R. Sabuncu\inst{1, 4}}
\authorrunning{A. Wang et al.}
%
\institute{School of Electrical and Computer Engineering, Cornell University \and Computer Science and Artificial Intelligence Lab at the Massachusetts Institute of Technology \and A.A. Martinos Center for Biomedical Imaging at the Massachusetts General Hospital \and Meinig School of Biomedical Engineering, Cornell University.}
\maketitle              
\begin{abstract}
Compressed Sensing MRI (CS-MRI) has shown promise in reconstructing under-sampled MR images, offering the potential to reduce scan times. Classical techniques minimize a regularized least-squares cost function using an expensive iterative optimization procedure. Recently, deep learning models have been developed that model the iterative nature of classical techniques by unrolling iterations in a neural network. While exhibiting superior performance, these methods require large quantities of ground-truth images and have shown to be non-robust to unseen data. In this paper, we explore a novel strategy to train an unrolled reconstruction network in an unsupervised fashion by adopting a loss function widely-used in classical optimization schemes. We demonstrate that this strategy achieves lower loss and is computationally cheap compared to classical optimization solvers while also exhibiting superior robustness compared to supervised models. Code is available at \url{https://github.com/alanqrwang/HQSNet}.

\end{abstract}

\section{Introduction}

Magnetic resonance (MR) imaging can be accelerated via under-sampling k-space -- a technique known as Compressed Sensing MRI (CS-MRI)~\cite{lustig}. This yields a well-studied ill-posed inverse problem. Classically, this problem is reduced to regularized regression, which is solved via an iterative optimization scheme, e.g.,~\cite{fista,admmboyd,primal_dual_chambolle,combettes2009proximal,daubechies2003iterative,Ye_2019}, on each collected measurement set. The limitations of this instance-based optimization approach are well-known; solutions are heavily influenced by the choice of regularization function and can lack in high-frequency detail. Furthermore, they are often time-consuming to compute.

There has been a recent surge in deep learning methods for CS-MRI, which promise superior performance and computational efficiency. 
To date, these methods have largely relied on fully-sampled data, which are under-sampled retrospectively to train the neural network model. 
The primary focus of this body of research has been the design of the neural network architecture. 
So-called ``unrolled architectures''~\cite{Aggarwal_2019,Hammernik_2017,liang2019deep,mardani2018neural,Schlemper_2017,Tezcan_2019,yang2017admmnet} which inject the MR-specific forward model into the network architecture have been shown to outperform more general-purpose, black-box~\cite{lee_deepresiduallearning,liang2019deep,wang_deeplearning} models and aforementioned classical methods. 

While these methods exhibit state-of-the-art reconstruction performance, a major limitation of the supervised formulation is the necessity for a dataset of fully-sampled ground-truth images, which can be hard to obtain in the clinical setting. In addition, these models are known to exhibit poor robustness at test-time when subjected to noisy perturbations and adversarial attacks~\cite{antun2019instabilities}.

In this work, we present a novel approach for performing MR reconstruction that combines the robustness of classical techniques with the performance of unrolled architectures. 
Specifically, we implement an unrolled neural network architecture that is trained to minimize a classical loss function, which does not rely on fully-sampled data. 
This is an ``amortized optimization'' of the classical loss, and we refer to our model as ``unsupervised''. 
We show that our unsupervised model can be more robust than its supervised counterpart under noisy scenarios. Additionally, we demonstrate that not only can we replace an expensive iterative optimization procedure with a simple forward pass of a neural network, but also that this method can outperform classical methods even when trained to minimize the same loss.

\section{Background}
In the CS-MRI formulation, fully-sampled MR images are assumed to be transformed into under-sampled $k$-space measurements by the forward model: 
\begin{equation}
    y = \mathcal{F}_\Omega x,
\end{equation}
where $x \in \mathbb{C}^N$ is the unobserved fully-sampled image, $y \in \mathbb{C}^M$ is the under-sampled $k$-space measurement vector\footnote{In this paper, we assume a single coil acquisition.}, $M < N$, and $\mathcal{F}_\Omega$ denotes the under-sampled Fourier operator with $\Omega$ indicating the index set over which the $k$-space measurements are sampled.
For each instance $y$, classical methods solve the ill-posed inverse problem via an optimization of the form:
\begin{equation}
    \argmin_x \norm{\mathcal{F}_\Omega x - y}^2_2 + \mathcal{R}(x),
    \label{eq:classical}
\end{equation}
where $\mathcal{R}(x)$ denotes a regularization loss term. The regularization term is often carefully engineered to restrict the solutions to the space of desirable images. Common choices include sparsity-inducing norms of wavelet coefficients~\cite{waveletnorm}, total variation~\cite{hdtv,1992PhyD...60..259R}, and their combinations~\cite{lustig,Ravishankar2020}. The first term of Eq.~\eqref{eq:classical}, called the data consistency term, quantifies the agreement between the measurement vector $y$ and reconstruction $x$.

Half-quadratic splitting (HQS) \cite{hqs_geman,hqs_analysis} solves Eq.~\eqref{eq:classical} by decoupling the minimization of the two competing terms using an auxiliary variable $z$
and an alternating minimization strategy over iterations $k\in \mathbb{N}$:
\begin{subequations} \label{eq.alternate-minimize}
\begin{align}
    z_{k} &= \argmin_z \mathcal{R}(z) + \lambda\norm{z - x_{k}}_2^2,\label{eq:z-minimization} \\
    x_{k+1} &= \argmin_x \norm{\mathcal{F}_\Omega x - y}^2_2 + \lambda\norm{z_{k} - x}_2^2,\label{eq:x-minimization}
\end{align}
\end{subequations}
where $\lambda \geq 0$ is a hyper-parameter.
Eq.~\eqref{eq:x-minimization} has closed-form solution in $k$-space at the sampling location $m$ given by:
\begin{equation}
    \hat{x}_{k+1}[m] = \begin{cases}
        \frac{y[m] + \lambda \hat{z}_{k} [m]}{1 + \lambda}, & \text{if  } m \in \Omega \\
        \hat{z}_{k} [m], &\text{else}
        \label{eq:closedform}
    \end{cases}
\end{equation}
for all $k$, where $\hat{x}_{k+1}$ and $\hat{z}_{k}$ denote $x_{k+1}$ and $z_{k}$ in Fourier domain, respectively. The $z$-minimization of Eq.~\eqref{eq:z-minimization} is the proximal operator for the regularizer, which may be solved using (sub-)gradient descent for differentiable $\mathcal{R}$. In this paper, we view the proximal operator as a function, i.e. $z_{k} = g(x_{k+1})$, where $g$ is some neural network. HQS and its data-driven variants underlie algorithms in CS-MRI \cite{Aggarwal_2019,Schlemper_2017}, image super-resolution \cite{CHENG2020103193}, and image restoration \cite{denoising-prior}.

Supervised deep learning offers an alternative approach.
Given a dataset of fully-sampled images and (often retrospectively-created) under-sampled measurements $\mathcal{D} = \{(x_i, y_i)\}_{i=1}^N$, these methods learn the optimal parameters $\theta$ of a parameterized mapping $G_\theta : y_i \rightarrow x_i$ by minimizing:
\begin{equation}
    \argmin_\theta \frac{1}{N} \sum_{i=1}^N \mathcal{L}_{sup}(G_\theta(y_i), x_i),
    \label{eq:supervised}
\end{equation}
where $\mathcal{L}_{sup}$ is a loss function that quantifies the quality of reconstructions based on the fully-sampled $x$. 
This formulation obviates the need for the design of a regularization loss function.
The parameterized mapping is often a neural network model~\cite{loupe,lee_deepresiduallearning,liang2019deep,wang_deeplearning}. 
Recently, unrolled architectures that exploit knowledge of the forward model \cite{Aggarwal_2019,mardani2018neural,Schlemper_2017} have proven to be effective. These architectures implement layers that iteratively minimize the data consistency loss and remove aliasing artifacts by learning from fully-sampled data.
$K \in \mathbb{N}$ such blocks are concatenated and  
trained end-to-end to minimize the supervised loss of Eq.~\eqref{eq:supervised}.


\section{Proposed Method}

To remove the need for fully-sampled data, leverage the robustness of the classical optimization method, and incorporate the performance of deep learning models, we propose to use an unsupervised strategy with an unrolled reconstruction architecture, which we call HQS-Net. 
Let a parameterized mapping $G_\theta$ denote a neural network that maps under-sampled measurements to reconstructed images. We train this network to minimize over $\theta$:
%
\begin{equation}
    \mathcal{L}(y_i; \theta) = \frac{1}{N} \sum_{i=1}^N \left[\norm{\mathcal{F}_\Omega G_\theta(y_i) - y_i}^2_2 + 
    \mathcal{R}\left(G_\theta(y_i)\right)\right].
    \label{eq:unsupervised}
\end{equation}
This model can be viewed as an amortization of the instance-specific optimization of Eq.~(\ref{eq:classical}), via a neural network $G_\theta$~\cite{Balakrishnan_2019,cremer2018inference,Gershman2014AmortizedII,marino2018iterative,shu2018amortized}. 

Amortized optimization provides several advantages over classical solutions. First, at test-time, it replaces an expensive iterative optimization procedure with a simple forward pass of a neural network. 
Second, since the function $G_\theta$ is tasked with estimating the reconstruction for any viable input measurement vector $y$ and not just a single instance, amortized optimization has been shown to act as a natural regularizer for the optimization problem~\cite{Balakrishnan_2019,shu2018amortized}.

\subsection{Model Architecture}
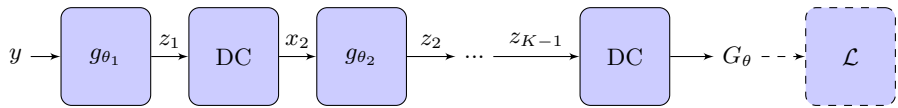
\begin{figure}[t]
\centering
\begin{tikzpicture}[node distance = 2cm, auto]
    \node [blank] (obs) {$y$};
	 \node [rblock, right of=obs, node distance=1.2cm] (iter1) {$g_{\theta_1}$};
	 \node [rblock, right of=iter1, node distance=1.7cm] (iter2) {DC};
	 \node [rblock, right of=iter2, node distance=1.7cm] (iter3) {$g_{\theta_2}$};
	 \node [blank, right of=iter3, node distance=1.5cm] (ellipsis) {...};
	 \node [rblock, right of=ellipsis, node distance=2cm] (iterinf) {DC};

    \node [blank, right of=iterinf, node distance=1.5cm] (end) {$G_\theta$};
    \node [dblock, right of=end, node distance=1.5cm] (loss) {$\mathcal{L}$};
	 \path [line] (obs) -- (iter1);
	 \path [line] (iter1) --  node [above] {$z_1$}(iter2);
	 \path [line] (iter2) --  node [above] {$x_2$} (iter3);
	 \path [line] (iter3) --  node [above] {$z_2$} (ellipsis);
	 \path [line] (ellipsis) -- node [above] {$z_{K-1}$} (iterinf);
    \path [line] (iterinf) --(end);    
    \path [line, dashed] (end) -- (loss);
\end{tikzpicture}
\caption{Proposed architecture. CNN and DC layers are unrolled $K$ times as a deep network, and the final output $G_\theta = x_K$ is encouraged to minimize $\mathcal{L}$ defined in Eq.~(\ref{eq:unsupervised}). $\mathcal{L}$ does not see fully-sampled data $x$.}
\label{fig:model-arch}
\end{figure}

Similar to instance-based iterative procedures like HQS and supervised unrolled architectures such as~\cite{Aggarwal_2019,Schlemper_2017}, HQS-Net decouples the minimization of the data consistency term and regularization term in Eq.~(\ref{eq:unsupervised}). 
Specifically, in each iteration block, the network explicitly enforces data consistency preceded by a convolutional block $g_{\theta_k}$ that learns an iteration-specific regularization.
Thus, we obtain an alternating minimization analogous to Eq.~\eqref{eq.alternate-minimize}:
\begin{subequations}
\begin{align}
    z_k &= g_{\theta_k}\left(x_k\right), \label{eq:regularizer-cnn} \\
    \hat{x}_{k+1}[m] &= \begin{cases}
        \frac{y[m] + \lambda \hat{z}_{k} [m]}{1 + \lambda}, & \text{if  } m \in \Omega \\
        \hat{z}_{k} [m], &\text{else}
        \label{eq:data-consistency}
    \end{cases}
\end{align}
\end{subequations}
where $x_1 = \mathcal{F}_\Omega^H y$ (i.e. the zero-filled reconstruction). Eq.~\eqref{eq:data-consistency} is implemented as a data-consistency layer (DC) within the network\footnote{For forward models that do not permit an analytical solution of Eq.~(\ref{eq:data-consistency}) (e.g. multi-coil MRI), one can replace the data-consistency layer with an iterative optimization scheme (e.g. conjugate gradient as in~\cite{Aggarwal_2019}). In addition, the iteration-specific weights $\theta_k$ in Eq.~(\ref{eq:regularizer-cnn}) can be replaced by a shared set of weights $\theta$, which enforces the model to learn a global regularization prior for all iterations.}.
The unrolled network concatenates $K \in \mathbb{N}$ of these $g_{\theta_k}$ and DC blocks, as shown in Fig.~\ref{fig:model-arch}.

\section{Experiments}

In our experiments, we used three different MRI datasets: T1-weighted axial brain, T2-weighted axial brain, and PD-weighted coronal knee scans. 
We applied retrospective down-sampling with $4$-fold and $8$-fold acceleration sub-sampling masks generated using a Poisson-disk variable-density sampling strategy~\cite{Chauffert_2013,Geethanath2013CompressedSM,lustig}. We use $2$nd-order and $3$rd-order polynomial densities for the $4$-fold and $8$-fold masks, respectively. All training and testing experiments in this paper were performed on a machine equipped with an Intel Xeon Gold 6126 processor and an NVIDIA Titan Xp GPU.

\subsubsection{Data.}

T1-weighted brain scans were obtained from~\cite{Dalca_2018_CVPR}, T2-weighted brain scans were obtained from the IXI dataset\footnote{https://brain-development.org/ixi-dataset}, and PD-weighted knee scans were obtained from the fastMRI NYU dataset~\cite{zbontar2018fastmri}. All images were intensity-normalized to the range $[0, 1]$ and cropped and re-sampled to a pixel grid of size $256 \times 256$. Dataset sizes consisted of $2000$, $500$, and $1000$ slices for training, validation, and testing, respectively.

\begin{figure}[t!]
\centering
\includegraphics[scale=0.5]{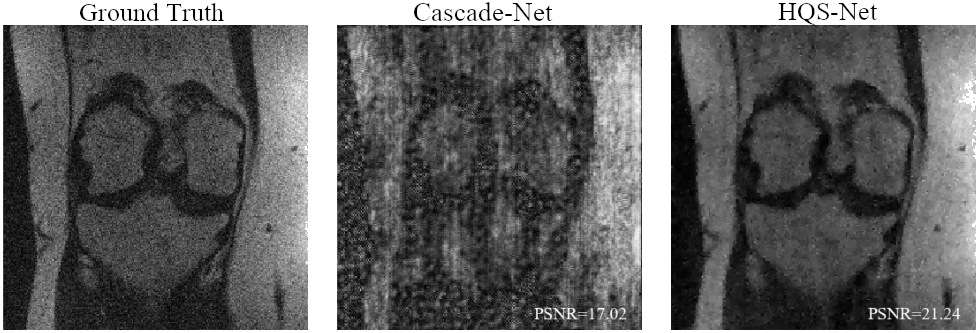}
\caption{Example of typical reconstructions of PD-weighted knee for $8$-fold acceleration under additive white Gaussian spatial noise, where noise standard deviation $\sigma=0.1$.} 
\label{fig:representative-slice-noise}
\end{figure}
\subsubsection{Comparison Models.} 

We compared the proposed HQS-Net against HQS and Cascade-Net~\cite{Schlemper_2017}, a supervised upper-bound baseline model. All models were implemented in Pytorch.

HQS minimizes the instance-based loss in Eq.~\eqref{eq:classical} using the alternating algorithm in Eq.~\eqref{eq.alternate-minimize}, where the $z$-minimization is performed using gradient descent. All iterative procedures were run until convergence within a specified tolerance. In choosing $\mathcal{R}(x)$, we followed the literature~\cite{lustig,Ravishankar2020} and let
\begin{equation}
\label{eq:reg-definition}
    \mathcal{R}(x) = \alpha TV(x) + \beta \norm{Wx}_1,
\end{equation}
where $TV$ denotes total variation, $W$ denotes the discrete wavelet transform operator, $\alpha, \beta > 0$ are weighting coefficients, and $\norm{\cdot}_1$ denotes the $\ell_1$ norm.

Cascade-Net is trained to minimize Eq.~\eqref{eq:supervised} using an $\ell_2$ loss and with an identical model architecture as HQS-Net.
Thus, Cascade-Net requires access to fully-sampled training data, whereas HQS and HQS-Net do not.


For $g_{\theta_k}$, a $5$-layer model was used with channel size $64$ at each layer. Each layer consists of convolution followed by a ReLU activation function. We used a residual learning strategy which adds the zero-filled input to the output of the CNN. The overall architecture $G_\theta$ is unrolled such that $K=25$. For training, Adam optimization~\cite{kingma2014adam} was used with a learning rate of 0.001 and batch size of 8. In experiments, we set $\lambda=1.8$, $\alpha = 0.005$, and $\beta = 0.002$, which were optimized using a Bayesian Optimization hyper-parameter tuning package~\cite{hyperopt}.

\begin{figure}[t]
\centering
\includegraphics[scale=0.3]{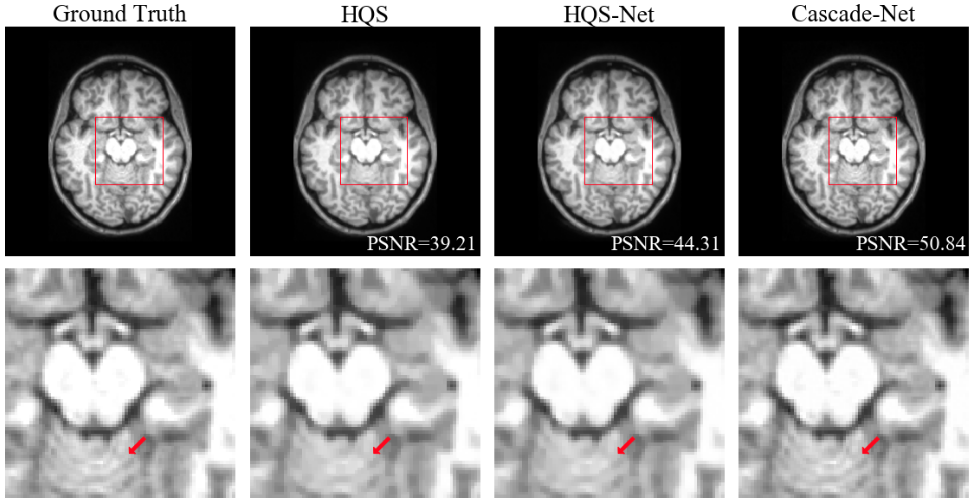}
\caption{Example reconstructions of a T1-weighted axial brain slice for $4$-fold acceleration. Although classical and unsupervised methods minimize the same loss, the unsupervised model is able to retain more high-frequency detail.} 
\label{fig:representative-slice-no-noise}
\end{figure}

\subsubsection{Evaluation Metrics.} Reconstructions were evaluated against ground-truth images on peak signal-to-noise ratio (PSNR), structural similarity index (SSIM)~\cite{ssim}, and high-frequency error norm (HFEN)~\cite{bresler-dictionarylearning}. The \textit{relative} value (e.g., relative PSNR) for a given reconstruction was computed by subtracting the corresponding metric value for the zero-filled reconstruction. 

\subsection{Results}
\subsubsection{Runtime and Loss Analysis.}
Table \ref{tab:runtime} shows the average runtime and average loss value (defined in Eq.~\eqref{eq:classical} and \eqref{eq:reg-definition}) achieved by both HQS and HQS-Net on the test set. Since inference for HQS-Net equates to a forward pass through the trained network, HQS-Net is several orders of magnitude faster while achieving superior loss compared to HQS.

\begin{table}
\caption{Inference runtime and loss. Lower is better. Mean $\pm$ standard deviation across test cases. 4-fold acceleration.}
\begin{tabular*}{\textwidth}{@{\extracolsep{\fill}}clcccccccc}
\hline
 \textit{Dataset} & \textit{Method} & \textit{Inference time (sec)}  & \textit{Loss} \\ \hline
 \multirow{2}{*}{T1 Brain} & HQS    & 483$\pm$111  &17.34$\pm$2.94 \\
& HQS-Net & 0.241$\pm$0.013  & 17.20$\pm$2.99  \\
\hline
 \multirow{2}{*}{T2 Brain} & HQS    & 380$\pm$72 & 19.93$\pm$4.20\\
& HQS-Net & 0.246$\pm$0.023  & 19.46$\pm$4.13 \\
\hline
  \multirow{2}{*}{PD Knee} & HQS    & 366$\pm$129 & 25.28$\pm$10.93\\
& HQS-Net & 0.251$\pm$0.013 & 24.61$\pm$10.58\\
\hline
\end{tabular*}
\label{tab:runtime}
\end{table}

\subsubsection{Robustness Against Additive Noise.}
\begin{figure}[t]
\centering
\includegraphics[width=\textwidth]{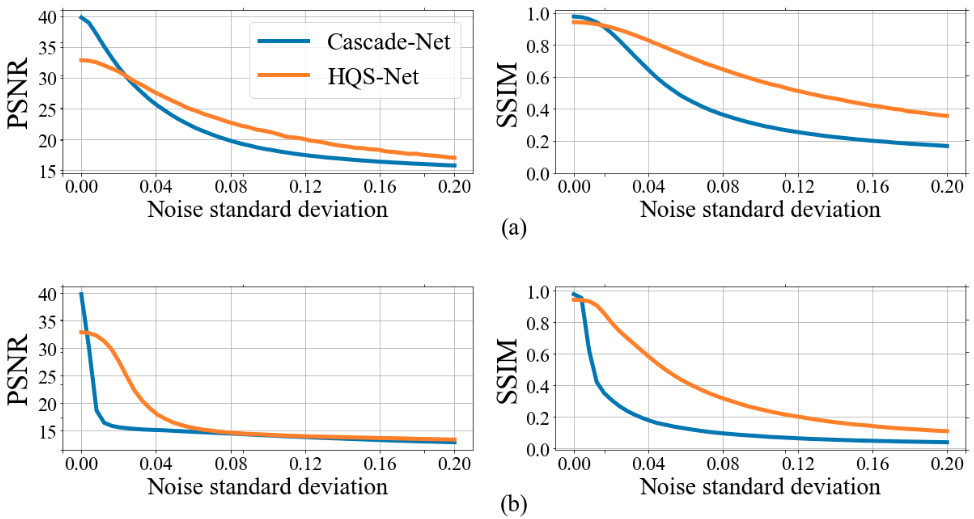}
\caption{Average reconstruction performance vs. noise standard deviation for $8$-fold acceleration, evaluated on PSNR and SSIM. (a) shows performance under additive noise in image domain and (b) shows performance under additive noise in $k$-space.} 
\label{fig:noise-curve-8fold}
\end{figure}

Since HQS-Net does not see fully-sampled data and is trained to minimize a robust classical loss, we expect it to exhibit better performance under circumstances of unseen data and/or noise compared to supervised models. To test this, we artificially inject additive Gaussian noise in both image space and $k$-space on the test set. Fig.~\ref{fig:noise-curve-8fold} shows a plot of reconstruction quality versus noise variance for $8$-fold acceleration. A visual example of the failure of supervised models to perform well under noisy conditions is shown in Fig.~\ref{fig:representative-slice-noise}.

\subsubsection{Comparison Models.}
\begin{figure}
\centering
\includegraphics[scale=0.31]{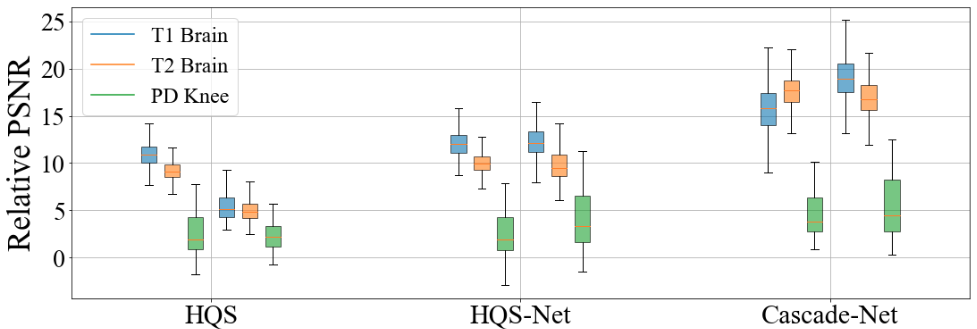}
\caption{Reconstruction performance for comparison models of all datasets evaluated on relative PSNR for 4-fold (left) and 8-fold (right) acceleration rates.} 
\label{fig:classic-sup-compare-all}
\end{figure}

Fig.~\ref{fig:classic-sup-compare-all} shows reconstruction performance of all three methods across three datasets. HQS-Net is comparable, if not superior (particularly at high acceleration rates), to the instance-based HQS solver despite the fact that they optimize the same loss function. This may be attributed to the network being trained across many samples, such that it is able to leverage commonalities in structure and detail across the entire training set. Fig.~\ref{fig:representative-slice-no-noise} highlights the learning of high-frequency detail.

\section{Conclusion}
We explored a novel unsupervised MR reconstruction method that performs an amortized optimization of the classical loss formulation for CS-MRI, thus eliminating the need for fully-sampled ground-truth data. We show that our method is more robust to noise as compared to supervised methods that have the same network architecture and is computationally cheaper than classical solvers that minimize the same loss. While our experiments focused on MRI, the method is broadly applicable to other imaging modalities and can be improved with more expressive networks and/or regularization functions. \blfootnote{Acknowledgements: This research was funded by NIH grants R01LM012719, R01AG053949; and, NSF CAREER 1748377, and NSF NeuroNex Grant1707312.}

\begin{table}
\caption{Relative Performance. Higher is better. Mean $\pm$ standard deviation across test cases. 4-fold acceleration.}
\begin{tabular*}{\textwidth}{@{\extracolsep{\fill}}clccccccccc}
\hline
          \textit{Dataset} & \textit{Method} & \textit{PSNR}  & \textit{SSIM}  & \textit{Negative HFEN} \\ \hline
 \multirow{2}{*}{T1 Brain} & HQS          & 10.88$\pm$1.256          & 0.401$\pm$0.052          & 0.154$\pm$0.016  \\
                           & HQS-Net      & 12.10$\pm$1.431 & 0.409$\pm$0.052 & 0.157$\pm$0.016  \\
                           & Cascade-Net  & 15.72$\pm$2.489          & 0.416$\pm$0.051          & 0.180$\pm$0.016   \\
\hline
 \multirow{2}{*}{T2 Brain} & HQS          & 9.266$\pm$0.961          & 0.369$\pm$0.057          & 0.172$\pm$0.014 \\
                           & HQS-Net      & 10.00$\pm$1.055 & 0.382$\pm$0.057 & 0.177$\pm$0.015 \\
                           & Cascade-Net  & 17.64$\pm$1.910          & 0.399$\pm$0.056          & 0.209$\pm$0.017 \\
\hline
  \multirow{2}{*}{PD Knee} & HQS          & 2.387$\pm$2.120          & 0.016$\pm$0.021          & 0.113$\pm$0.057 \\
                           & HQS-Net      & 2.472$\pm$2.117 & 0.018$\pm$0.021 & 0.120$\pm$0.054 \\
                           & Cascade-Net  & 4.419$\pm$2.081          & 0.086$\pm$0.023          & 0.179$\pm$0.048\\
\hline
\end{tabular*}
\vspace{8pt}
\caption{Relative Performance. 8-fold acceleration}
\begin{tabular*}{\textwidth}{@{\extracolsep{\fill}}clccccccccc}
\hline
 \textit{Dataset} & \textit{Method} & \textit{PSNR}  & \textit{SSIM}  & \textit{Negative HFEN} \\ \hline
 \multirow{2}{*}{T1 Brain} & HQS          & 5.403$\pm$1.500          & 0.161$\pm$0.055          & 0.303$\pm$0.045  \\
                           & HQS-Net      & 12.31$\pm$1.814 & 0.566$\pm$0.052 & 0.436$\pm$0.035  \\
                           & Cascade-Net   & 15.38$\pm$2.570          & 0.586$\pm$0.049          & 0.531$\pm$0.025   \\
\hline
 \multirow{2}{*}{T2 Brain} & HQS          & 5.261$\pm$1.520          & 0.211$\pm$0.050          & 0.310$\pm$0.055 \\
                           & HQS-Net      & 9.880$\pm$1.839 & 0.553$\pm$0.052 & 0.457$\pm$0.031 \\
                           & Cascade-Net    & 16.92$\pm$1.898          & 0.589$\pm$0.044          & 0.576$\pm$0.025 \\
\hline
  \multirow{2}{*}{PD Knee} & HQS          & 2.109$\pm$1.277          & 0.051$\pm$0.040          & 0.199$\pm$0.094 \\
                           & HQS-Net      & 4.019$\pm$2.782 & 0.085$\pm$0.063 & 0.252$\pm$0.101 \\
                           & Cascade-Net    & 5.393$\pm$3.043          & 0.156$\pm$0.038          & 0.347$\pm$0.103\\
\hline
\end{tabular*}
\end{table}
%
%
%
\clearpage

\end{document}